\documentstyle[twocolumn,aps,psfig]{revtex}

\begin{document}
\draft
\flushbottom
\twocolumn[
\hsize\textwidth\columnwidth\hsize\csname @twocolumnfalse\endcsname

\title{ Giant Enhancement of Surface Second Harmonic Generation in
BaTiO$_3$ due to Photorefractive Surface Wave Excitation. }
\author{Igor I. Smolyaninov, Chi H. Lee, Christopher C. Davis}
\address{ Electrical Engineering Department \\
University of Maryland, College Park,\\
MD 20742}
\date{\today}
\maketitle
\tightenlines
\widetext
\advance\leftskip by 57pt
\advance\rightskip by 57pt

\begin{abstract}
We report observation of strongly enhanced surface SHG in BaTiO$_3$ due to 
excitation of a photorefractive surface electromagnetic wave. Surface SH 
intensity may reach $10^{-2}$ of the incident fundamental light intensity. 
Angular, crystal orientation, and polarization dependencies of this SHG are 
presented. Possible applications of this effect in nonlinear surface 
spectroscopy are discussed. 


\end{abstract}

]
\narrowtext

\tightenlines

Linear surface electromagnetic waves (SEW) such as surface plasmons or surface 
polaritons \cite{1} play a very important role in such surface optical phenomena 
as surface enhanced Raman scattering, surface second harmonic generation (SHG), 
etc. They are extremely useful in applications such as chemical and biological 
sensing since the electromagnetic field of a SEW is strongly enhanced near the 
interface. A linear SEW may be excited at the interface between media with 
opposite signs for their dielectric constants $\epsilon $, such as at a metal-
vacuum interface. Another example is the interface between a vacuum and a 
dielectric that has a sharp absorption line. Such a dielectric has $\epsilon 
(\omega )<0$ for frequencies just above the absorption line. In both cases the 
SEW free propagation length in the visible range does not exceed a few 
micrometers because of high losses \cite{1}. This limits the potential 
advantages of using SEW in surface enhanced nonlinear optical studies and
sensor applications.

Recently, a new kind of nonlinear SEW called a photorefractive surface wave has 
been predicted \cite{2} and observed experimentally \cite{3} in BaTiO$_3$. This 
phenomenon is closely related to self-trapped optical beams (also known as 
spatial solitons) and self-bending beams observed in photorefractive crystals 
\cite{4}. A photorefractive SEW occurs when a beam self-bent towards the 
positive direction of the optical axis (the poling direction) undergoes
a cycle of deflections towards the face of the crystal and total internal 
reflections. The resulting nonlinear SEW is localized near the crystal-air 
interface with a penetration depth as small as 10 micrometers into the 
photorefractive crystal \cite{2}. This leads to a strong enhancement of the 
optical field near the interface that is common for all SEWs. On the other hand,
since BaTiO$_3$ is transparent in the visible range, the free propagation length 
of the photorefractive SEW along the surface is limited only by the size of the 
crystal. As a result, a very strong enhancement of all nonlinear surface optical 
phenomena (such as surface adsorbed molecular luminescence, Raman scattering, 
surface SHG, etc.) may be expected due to photorefractive SEW excitation. This 
effect may also be used in combination with further field enhancements
produced by surface topographical defects or by the probe tip of a scanning 
probe microscope (the field enhancement by a probe tip is discussed in the 
context of a feasibility study of molecular resolution fluorescence near-field 
microscopy using multi-photon absorption by Kawata $et$ $al.$ \cite{5}).

In this letter we report the first observation of strongly enhanced surface SHG 
due to the photorefractive SEW excitation in BaTiO$_3$. Surface SHG in BaTiO$_3$ 
is a very suitable phenomenon for demonstrating the potential of photorefractive 
SEWs in nonlinear surface optics. Phase-matched optical SHG (collinear or non-
collinear) is forbidden in the bulk of BaTiO$_3$ in the visible range because of 
strong dispersion: the refractive indices for ordinary and extraordinary
waves are 2.67 and 2.57 respectively at 400 nm, and 2.36 and 2.32 respectively 
at 800 nm light wavelength \cite{6}. The momentum conservation law can not be 
satisfied in the volume SHG process. Thus, in contrast to experiments done with 
KTP crystals \cite{7}, observation of the fundamental and phase-matched SHG 
fields mutually trapped in the volume spatial solitary 
wave is impossible in BaTiO$_3$. On the other hand, phase-matching conditions 
are modified for the surface SHG. Only the momentum component parallel to the 
interface must be conserved. This leads to an extremely strong phase-matched 
surface SHG when a photorefractive SEW is excited.

The surface nature of this SHG, in combination with strong photorefractivity of 
BaTiO$_3$ leads to quite peculiar angular and orientational behavior of the SH 
light intensity. We believe that our observations performed in a well controlled 
geometry may clarify a lot of questions concerning still unclear "anomalous" SHG 
in BaTiO$_3$ reported recently by a number of groups \cite{8,9}. This is 
especially important for the rapidly developing field of ferroelectric oxide 
film growth and characterization. Much recent effort in this field is caused by 
the applications of these films in nonvolatile ferroelectric random-access 
memory (NVFRAM) and dynamic random-access memory (DRAM) devices \cite{10}. SHG 
has been used to determine the crystallographic orientation and the degree of 
poling of these films (in particular, BaTiO$_3$ films in \cite{9}).

\begin{figure}[tbp]
\centerline{
\psfig{figure=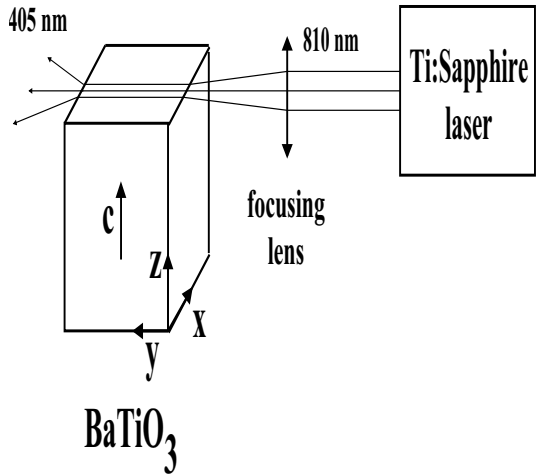,width=7.0cm,height=6.5cm,clip=}
}
\caption{ Schematic view of our experimental geometry.  }
\label{fig1}
\end{figure}

Our experimental geometry is shown in Fig.1. The iron doped crystal of BaTiO$_3$ 
from Sanders Inc. used in the experiments is a $8 mm \times 8 mm \times 8 mm$ 
cube which was cut with 2 opposing faces perpendicular to the c axis and poled 
along the c axis. The crystal was mounted on a three-axis translational and 
rotational stage. Linearly polarized light from Ti:Sapphire laser operating at a 
wavelength of 810 nm (repetition rate 76 MHz, 100 fs pulse duration, 30 mW 
average power) was focused onto the top edge of the crystal. The
direction of the beam was parallel to the top face of the crystal. 

\begin{figure}[tbp]
\centerline{
\psfig{figure=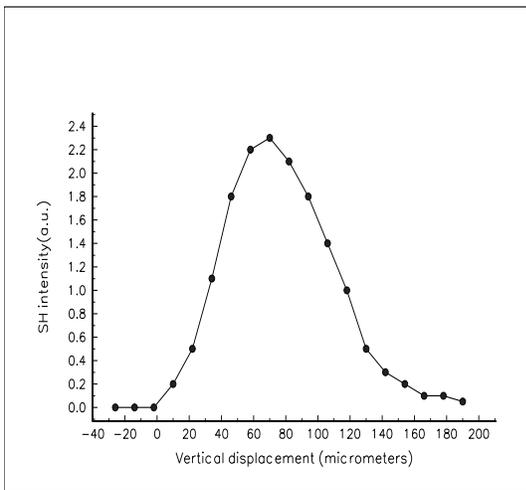,width=7.0cm,height=6.5cm,clip=}
}
\caption{ SH intensity as a function of the vertical position $z$ of the 
crystal. Large positive
$z$ corresponds to the fundamental light passing through the crystal. 
Approximate position
of the focal spot of the lens is shown by the arrow.  }
\label{fig2a}
\end{figure}

\begin{figure}[tbp]
\centerline{
\psfig{figure=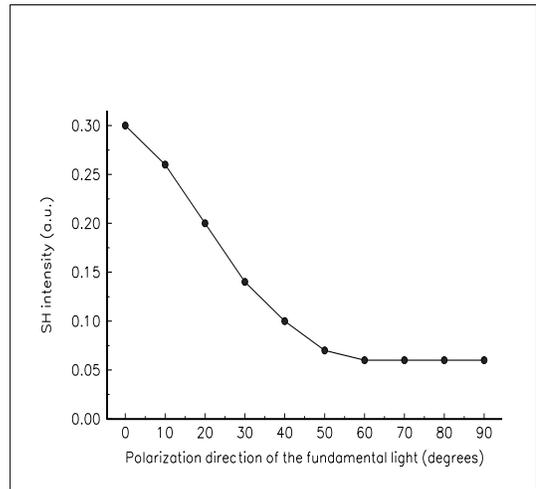,width=7.0cm,height=6.5cm,clip=}
}             
\caption{ SH intensity as a function of the polarization direction of the 
linearly polarized fundamental light, 90 degrees corresponds to the 
polarization direction parallel to the optical axis of the BaTiO$_3$ crystal (z 
direction).}
\label{fig2b}
\end{figure}    

 As the vertical position of the crystal was scanned, very bright second 
harmonic light emission coming from the top face of the crystal was observed. 
The brightness of SH emission was very 
sensitive to the vertical position of the crystal as it is evident from 
Fig.2, which was obtained using a 
focusing lens with a focal length of 60 mm. No SH emission had been detected 
when the fundamental light passed through the crystal in any direction far from 
the top face. This is consistent with the fact that phase-matched SHG is 
prohibited in the bulk of BaTiO$_3$ crystal.
Also, no comparable SHG has been detected from the other five faces of the same 
crystal. The full width at half maximum of the SH peak in Fig.2 is equal to 
80 $\mu $m. 

The SH emission appeared to be localized in the plane of the top face of the 
crystal coming out of the top face within a wide (almost 180 degrees) angle. 
This peculiar spatial distribution of SHG is illustrated in Fig.4 which shows a 
pattern of SH illumination on the screen placed just behind the crystal. The SH 
light is vertically polarized. Its intensity is 
proportional to the square of the fundamental light intensity and depends 
strongly on the polarization state of the linearly polarized fundamental light 
(Fig.3). The SH intensity is approximately 6 times stronger in the case of 
fundamental light linearly polarized perpendicular to the optical axis of the 
BaTiO$_3$ crystal. 
   
\begin{figure}[tbp]
\centerline{
\psfig{figure=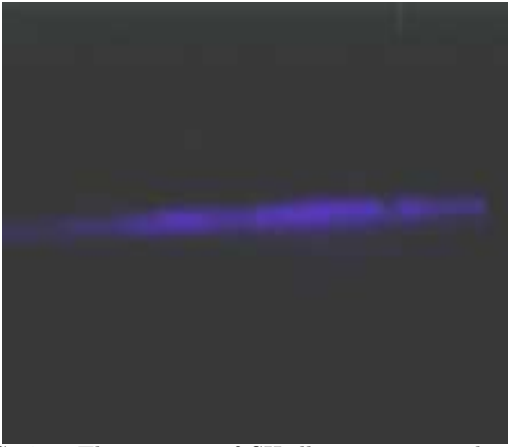,width=7.0cm,height=6.5cm,clip=}
}     
\caption{ The pattern of SH illumination produced on a screen placed just behind 
the crystal. A band pass filter is used to cut off the illumination by 
fundamental light.    }
\label{fig3}
\end{figure} 

All features of the observed SH emission are consistent with the proposed 
surface SHG enhancement due to the photorefractive surface wave excitation in 
BaTiO$_3$. The crystal cut allows photorefractive SEW propagation only on the 
top face of the crystal (only near this face can
photorefractive self-bending and total internal reflection compete with each 
other). This is consistent with the observation of SHG only from the top face of 
the crystal. At the same time, the full width at half maximum of the vertical 
position dependence of SHG in Fig.2 corresponds to the 
observed range of photorefractive SEW launching in BaTiO$_3$ \cite{3}. 

An explanation of the angular distribution of the observed SHG needs a more 
detailed analysis. Nonlinear SEW solutions of the Maxwell equations in a 
photorefractive medium may be found as follows \cite{2}. It is assumed that a 
SEW written as $E(z,y)=E(z)exp(-iky)$ induces a dielectric
constant changes of the form $\epsilon (z)=\epsilon +\delta \epsilon (z)$ due to 
the photorefractive effect (here $E(z)$ is supposed to be real, $\epsilon $ is 
the real dielectric constant of the medium in the absence of the wave, the 
photorefractive medium is supposed to be optically isotropic, and the field 
distribution is assumed to be independent of $x$. These 
simplifying assumptions may not be true in the most general case, but they allow 
us to illustrate the basic physics of the phenomenon). Substituting this into 
Maxwell's equations results in the following equation for E(z):

\begin{equation} 
(d^2/dz^2+(k_0^2-k^2)+k_0^2\delta \epsilon (z)/\epsilon )E(z)=0
\end{equation}

where $k_0=\omega (\epsilon \epsilon _0\mu _0)^{1/2}$ is the wave number of the 
light in a linear medium with the same unperturbed dielectric constant $\epsilon 
$. Assuming the diffusion mechanism for nonlinearity [11], the distribution of 
$\delta \epsilon (z)$ may be related to E(z) through the space-charge electric 
field $E_{sc}$:

\begin{equation}
E_{sc}=-(k_BT/e)(dI(z)/dz)/I(z) 
\end{equation}

where I(z) is the intensity of light. The space-charge electric field induces 
refractive index changes via the electro-optic effect \cite{11}:

\begin{equation}
\delta \epsilon (z)=2n^4r(k_BT/e)(dE(z)/dz)/E(z) 
\end{equation}

where $r$ is the linear electro-optic coefficient. Thus, we obtain the wave 
equation in the form

\begin{equation}
((d^2/dz^2)/k_0^2+(2\gamma d/dz)/k_0-2(k-k_0)/k_0)E(z)=0
\end{equation}

where $\gamma =k_0n^2r(k_BT/e)$. This equation has a solution exponentially 
decaying into the photorefractive medium:

\begin{equation}
E(z)=exp(-\gamma k_0z)cos((2(k-k_0)k_0)^{1/2}z+\phi)
\end{equation}

In the case of BaTiO$_3$ the photorefractive SEW field penetration depth is 
$d_z=(\gamma k_0)^{-1}\sim 10\mu m$ \cite{2}. Since the resulting equation (4) 
is linear, this penetration depth does not depend on the field intensity.

Real laser beams have finite width in $x$, but if the width of the beam is much 
larger than $d_z$ the simplified theory described above should be applicable. 
Thus, a real life photorefractive SEW must have a localization length in the $x$ 
direction ($d_x$) much larger than $d_z$. Indeed, this is evident from the 
profile of the photorefractive SEW in BaTiO$_3$ measured in \cite{3}, where 
$d_x \sim 500 \mu m$. Unlike $d_z$, $d_x$ may depend on the intensity of SEW.

A very fruitful approach to all phenomena related to spatial solitons is the 
representation of solitons as linear waves propagating in the self-induced 
optical waveguides \cite{12,13}. Let us follow this way of thinking and consider 
fundamental light propagating in a surface waveguide
with a rectangular $d_z \times d_x$ profile such as $d_x>>d_z \sim 10 \mu m$ 
(Fig.5). The phase matching conditions for SHG in the surface waveguide are 
further modified. Only the momentum component parallel to the waveguide 
direction must be conserved. At the same time,
such a waveguide is highly multimode with many optical modes corresponding to 
geometrical optics rays propagating in a zig-zag manner parallel to the surface. 
Thus, there will always be a SH waveguide mode phase-matched with the 
fundamental light. Upon resonant excitation, such a mode will be 
coupled to many other modes existing in the waveguide at the same frequency. As 
a result, when the SH light reaches the edge of the crystal, it leaves the 
surface waveguide in a ray fan parallel to the face of the crystal. This results 
in the pattern of SH illumination detected in Fig.4.

\begin{figure}[tbp]
\centerline{
\psfig{figure=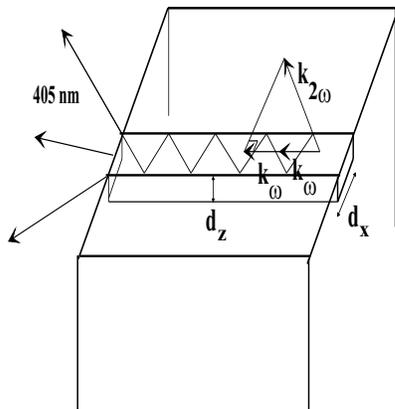,width=7.0cm,height=6.5cm,clip=}
}     
\caption{Phase-matched SHG in the self-induced surface waveguide.}
\label{fig4a}
\end{figure} 

\begin{figure}[tbp]
\centerline{
\psfig{figure=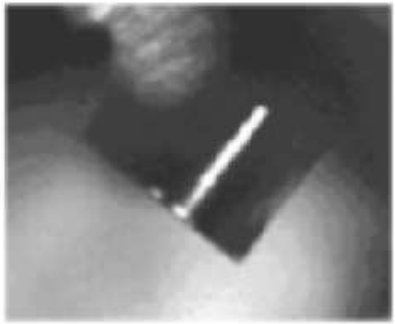,width=7.0cm,height=6.5cm,clip=}
}     
\caption{ A picture of the top face of the crystal taken through the bandpass 
filter which cuts off the fundamental illumination light. Simultaneous weak 
illumination with the white light from the flashlight allows to show the edges 
of the BaTiO$_3$ crystal. The focused fundamental beam was coming from the top 
right corner of the image. The surface beam of SH light trapped by the 
propagating photorefractive SEW is clearly visible.}
\label{fig4b}
\end{figure} 

We have performed further experiments to confirm the physical picture of the 
phenomenon described above. Using a 3.6-mm focal length $40 \times $ microscope 
objective as a focusing lens in Fig.1, we have observed much brighter surface SH 
emission with an average power on the order 
of some tens of microwatts. No damage to the surface of the BaTiO$_3$ crystal 
was apparent after this experiment. Thus, as much as $10^{-2}$ of the 
fundamental light power had been converted
into SH emission. This is a much higher conversion efficiency than is usually 
observed in surface SHG experiments: simple calculations show that as much as 
$10^7$ second harmonic photons per 1 nJ laser pulse have been generated. This is 
three orders of magnitude higher than the conversion efficiency observed in the 
case of surface SHG enhanced by a surface plasma wave \cite{14}.
It is comparable (just an order of magnitude smaller) with the best SH 
conversion efficiencies obtained in any other experimental geometry.
At such a high power self-pumped phase conjugation builds up with the laser
power dependent time constant on the order of a few hundreds of milliseconds,
causing instability of the Ti:Sapphire laser. So, optical isolation is necessary 
for SHG to be stable (which itself exhibits a similar characteristic buildup 
time following sudden changes in optical alighnment).  

A picture of the top face of the crystal taken through the bandpass filter 
(which cuts off the fundamental excitation light) is shown in Fig.6. 
Simultaneous weak illumination with the
white light from a flashlight allows us to show the edges of the BaTiO$_3$ 
crystal. The focused fundamental beam was coming from the top right corner of 
the image. The surface beam of SH light 
trapped by the propagating photorefractive SEW is clearly visible. The beam 
starts from the focal point of the microscope objective. It is possible to see 
the SHG because of surface scattering. The width of the self induced surface 
waveguide is estimated to be on the order of 400 $\mu $m.

The polarization properties of SH emission mostly reflect the properties of the 
second harmonic polarization tensor $d_{ijk}^{(2)}$ of BaTiO$_3$, which relates 
SH and the fundamental excitation light 
($P_i^{(2)}=\epsilon _0d_{ijk}^{(2)}E_j^{(1)}E_k^{(1)}$). For a surface SH 
process in the geometry of our experiment only two non-zero components of 
$d_{ijk}^{(2)}$ are available: $d_{zxx}^{(2)}=-18.8 \times 10^{-12} m/V$ and 
$d_{zzz}^{(2)}=-7.1 \times 10^{-12} m/V$ \cite{15}.
Thus, the ratio of SH intensity for excitation with fundamental light linearly 
polarized in the $x$ (horizontal) and $z$ (vertical) directions should be 
$(d_{zxx}^{(2)}/d_{zzz}^{(2)})^2=7.0$ which
is very close to the ratio observed in the experiment (Fig.3).

At the same time, the photorefractive coupling constant $r$ is smaller in the 
case of vertically (ordinary) polarized light. This means that the 
photorefractive SEW field penetration depth
$d_z$ is bigger in this case and the fundamental excitation light spends more 
time away from the surface. According to the simple self-induced waveguide model 
described above, we should expect stronger attenuation of SHG under this 
circumstances. In order to account for this discrepancy
a complimentary point of view on the nature of SHG enhancement due to the 
photorefractive SEW excitation may be suggested. In the picture of a SEW as a 
beam undergoing a cycle of photorefractive deflections towards the face of the 
crystal and total internal reflections, a periodic self-induced modulation of 
the refractive index near the surface of BaTiO$_3$ crystal may be 
expected. Such a situation would closely resemble the geometry of SHG 
experiments with periodically poled nonlinear crystals such as lithium niobate, 
which show substantial enhancement of SHG efficiency (periodic poling creates 
periodic modulation of refractive index near the interface). 

In conclusion, we have observed strongly enhanced surface SHG in BaTiO$_3$ due 
to excitation of a photorefractive surface electromagnetic wave. The surface SH 
intensity may reach $10^{-2}$ of the incident fundamental light intensity. A 
physical picture of this SHG has been introduced that assumes phase-matched SHG 
in the self-induced surface waveguide. Peculiar angular, crystal 
orientation, and polarization dependencies of this SHG are presented. They may 
account for anomalous SHG in BaTiO$_3$ recently reported in the literature. The 
observed phenomenon may have many potential applications in nonlinear surface 
spectroscopy. Also, it may be considered as an example of guiding of light with 
light in a self-induced surface waveguide,
which may find applications in such emerging soliton related techniques as 
writing virtual photonic circuits \cite{16}, etc.

\noindent {\bf Acknowledgment}: We would like to acknowledge helpful discussions 
with Kyuman Cho and Saeed Pilevar.


\begin{references}
\bibitem{1} Surface Polaritons, edited by V.M. Agranovich and D.L. Mills (North-
Holland, Amsterdam, 1982); H. Raether, Surface Plasmons, Springer Tracts in 
Modern Physics Vol.111 (Springer, Berlin, 1988). 

\bibitem{2} G.S. Garcia Quirino, J.J. Sanchez-Mondragon, and S. Stepanov, 
Phys.Rev.A 51, 1571 (1995).

\bibitem{3} M. Cronin-Golomb, Optics Letters 20, 2075 (1995).

\bibitem{4} B. Crosignani, P. Di Porto, M. Segev, G. Salamo, and A. Yariv, 
Rivista del Nuovo Cimento 21, 1 (1998).

\bibitem{5} Y. Kawata, C. Xu, and W. Denk, J.Appl.Phys. 85, 1294 (1999).

\bibitem{6}	Handbook of Lasers with Selected Data on Optical Technology, edited 
by R.J. Pressley (CRC Press, Clevelend, Ohio, 1971), p.509.

\bibitem{7} W.E. Torruellas, Z. Wang, D.J. Hagan, E.W. VanStryland, G.I. 
Stegeman, L. Torner, and C.R. Menyuk, Phys.Rev.Lett. 74, 5036 (1995).

\bibitem{8} E.V. Bursian, V.G. Zalesskii, A.A.Luzhkov, and V.V. Maslov, JETP 
Letters 64, 270 (1996).

\bibitem{9} L.D. Rotter, D.L. Kaiser, and M.D. Vaudin, Appl.Phys.Lett. 68, 310 
(1996).

\bibitem{10} O. Auciello, J.F. Scott, and R. Ramesh, Physics Today N7, 22 
(1998).

\bibitem{11} M.P. Petrov, S.I. Stepanov, and A.V. Khomenko, Photorefractive 
Crystals in Coherent Optical Systems (Springer, Berlin, 1991).

\bibitem{12} Y.R. Shen, Science 276, 1520 (1997); 

\bibitem{13} A.W. Snyder and D.J. Mitchell, Science 276, 1538 (1997).

\bibitem{14} Y.R. Shen, The Principles of Nonlinear Optics (Wiley, New York, 
1984) pp.479-504.

\bibitem{15} F. Zernike and J.E. Midwinter, Applied Nonlinear Optics (Wiley, New 
York, 1973)

\bibitem{16} A.W. Snyder and F. Ladouceur, Optics and Photonics News, Vol.10, 
N2, 35 (1999).
\end{references}
\end{document}